\documentclass{cimento}

\newcommand{\xd}{\dot{x}}

\newcommand{\qm}{\frac{q}{m}}
\newcommand{\ca}[1]{{\cal {#1}}}
\usepackage{amsmath,amssymb,bm}
\usepackage{cite,epsfig,color}
\title{Discontinuous distributions in thermal plasmas}
\author{D.A.~Burton\from{ins:lancs}\from{ins:ci},
A.~Noble\from{ins:lancs}\from{ins:ci}, H.~Wen\from{ins:lancs}\from{ins:ci}}
\instlist{\inst{ins:lancs} Department of Physics, Lancaster
University, Lancaster, UK
\inst{ins:ci} The Cockcroft Institute of Accelerator Science and
Technology, Daresbury, UK
}
\PACSes{\PACSit{52.65.Ff}{Fokker-Planck and Vlasov equation}
\PACSit{52.35.Fp}{Electrostatic waves and oscillations}}
\begin{document}

\maketitle

\begin{abstract}
We develop a new method for describing the dynamics of
$3$-dimensional thermal plasmas.  Using a piecewise constant
$1$-particle distribution, we reduce the Vlasov equation to a
generalized Lorentz force equation for a family of vector fields
encoding the discontinuity. By applying this equation to longitudinal
electrostatic plasma oscillations, and coupling it to Maxwell's
equations, we obtain a limit on the magnitude of the electric field in
relativistic thermal plasma oscillations. We derive an upper bound on
the limit and discuss its applicability in a background magnetic field. 
\end{abstract}

\section{Introduction}
High-power lasers and plasmas may be
used to accelerate electrons by electric fields that are
orders of magnitude greater than those achievable using conventional
methods~\cite{tajima:1979}.
An intense laser pulse is used to drive a wave in a
plasma and, for sufficiently large fields, non-linearities lead to
collapse of the wave structure (`wave-breaking') due to
sufficiently large numbers of electrons becoming trapped in the
wave.

Hydrodynamic investigations of wave-breaking were first undertaken for cold
plasmas~\cite{akhiezer:1956, dawson:1959} and thermal effects were
later included in non-relativistic~\cite{coffey:1971} and relativistic
contexts~\cite{katsouleas:1988, rosenzweig:1988, schroeder:2005}
(see~\cite{trines:2006}
for a discussion of the numerous approaches). However, it is
clear that the value of the electric field at which the wave
breaks (the electric field's `wave-breaking limit')
is highly sensitive to the details of the hydrodynamic model.  

Plasmas dominated by collisions are described by a pressure tensor that does
not deviate far from isotropy, whereas an intense and ultrashort laser
pulse propagating through an underdense plasma will drive the plasma
anisotropically over typical acceleration timescales. Thus, it is
important to accommodate 3-dimensionality and allow for anisotropy
when investigating wave-breaking limits.

Our aim is to uncover the relationship between wave-breaking limits
and the shape of the $1$-particle distribution $f$. In general, the detailed structure of $f$ cannot be reconstructed from a
few low-order moments so we adopt a different approach based on a
particular class of piecewise constant $1$-particle
distributions. Our approach may be considered
as a multi-dimensional generalization of the 
1-dimensional relativistic `waterbag' model employed
in~\cite{katsouleas:1988} (for a discussion of the
relationship between our approach and~\cite{katsouleas:1988} see~\cite{us:2008}).

We employ the Einstein summation
convention throughout and units are used in which the speed of light
$c=1$ and the permittivity of the vacuum $\varepsilon_0=1$. Lowercase
Latin indices $a,b,c$ run over $0,1,2,3$.
\section{Vlasov-Maxwell system}
Our attention is focussed on plasmas evolving over
timescales during which the `discrete' nature (collisions) of the plasma
electrons can be neglected and the plasma ions can be prescribed as a
background. Such configurations are well described by
the covariant Vlasov-Maxwell system~\cite{degroot:1980, ehlers:1971} which, for the
purposes of this paper, is most usefully expressed in the language of exterior
calculus (see, for example,~\cite{burton:2003, benn:1987}). We will
now briefly summarize the particular formulation of the Vlasov-Maxwell
system employed here.

Let $(\ca{M},g)$ be a spacetime
with signature $(-,+,+,+)$ for the metric tensor $g$. Each
point $p\in\ca{M}$ is associated with a space $\ca{E}_p\subset T_p\ca{M}$
of future-directed unit normalized vectors on $\ca{M}$,
\begin{equation}
 \ca{E}_p = \{ (x(p),\xd) \in T_p\ca{M} : g_{ab}(x(p))\xd^a\xd^b =-1 \text{ and } \xd^0>0 \},
\end{equation}
where $g_{ab}$ are the components of the metric $g$ in a coordinate
system $(x^a)$ whose patch contains $p$ and $(x^a,\xd^b)$ are induced
coordinates on $T\ca{M}$. The total space $\ca{E}$ of the
bundle $(\ca{E},\Pi,\ca{M})$ is the union of $\ca{E}_p$ over
$p\in\ca{M}$ and $\Pi$ is the restriction to $\ca{E}$ of the canonical
projection on $T\ca{M}$.

Naturally induced tensors on $T\ca{M}$ include the dilation vector field $X$
\begin{equation}
X = \xd^a \partial_a^{\bm{V}},
\end{equation}
the vertical lift $\star 1^{\bm{V}}$ of the volume $4$-form $\star
1$ from $\ca{M}$ to $T\ca{M}$ and the horizontal $4$-form $\# 1$
\begin{equation}
\# 1 = \sqrt{|\text{det}\mathfrak{g}|}^{\bm{V}} dx^{0\bm{H}}\wedge
dx^{1\bm{H}} \wedge dx^{2\bm{H}} \wedge dx^{3\bm{H}}
\end{equation}
where $dx^{a\bm{H}}$ is the horizontal lift of $dx^a$ from $\ca{M}$
to $T\ca{M}$ (see Appendix~\ref{appendix:vertical_and_horizontal} for
further details) and $\mathfrak{g}=(g_{ab})$ is the matrix of
components of $g$.

The Vlasov-Maxwell system for $f$ (a
scalar field on $T\ca{M}$ whose restriction to $\ca{E}$ is the plasma electron
$1$-particle distribution) and the electromagnetic field $F$ may be written
\begin{align}
\label{LVlas}
& Lf \simeq 0,\\
\label{Max}
& dF=0, \qquad d\star F= q\star (\widetilde{N}_\text{ion} -
  \widetilde{N})
\end{align}
where $\simeq$ indicates equality on restriction by pullback from $T\ca{M}$
to $\ca{E}$ and
\begin{align}
\label{Liou}
&L = \xd^a (\partial_a^{\bm{H}} + {\bf f}^{\bm{V}}_a) \in
TT\ca{M},\\
\label{force}
&{\bf f}_a= -\qm F^b{}_a \partial_b \in T\ca{M}
\end{align}
with $m$ the mass and $q$ the charge of the electron ($q<0$) and
$F^a{}_b= g^{ac} F_{cb}$ the components of the electromagnetic
2-form $F=\frac{1}{2} F_{ab} dx^a \wedge dx^b$ on $\ca{M}$.
The metric dual of a vector $V$ is defined by
$\widetilde{V}(Y)=g(Y,V)$ for all vectors $Y$ on $\ca{M}$ and $\star$
is the Hodge map induced from the volume $4$-form $\star
1$ 
\begin{equation}
\star 1 = \sqrt{|\text{det}\mathfrak{g}|}\, dx^0\wedge
dx^1 \wedge dx^2 \wedge dx^3
\end{equation}
on $\ca{M}$. The components of the electron number $4$-current
$N=N^a(x(p))\partial_a$ at $p\in\ca{M}$ are given as an integral over the fibre
$\ca{E}_p=\Pi^{-1}(p)$ 
\begin{align}
\notag
N^a(x(p)) &= \int_{\Pi^{-1}(p)} \xd^a f\iota_X\# 1\\
\label{electron_current}
&= -\int_{\Pi^{-1}(p)} \xd^a f(x(p),\xd)
\frac{\sqrt{|\text{det}{\mathfrak{g}(x(p))}|}}{g_{0c}(x(p))\xd^c} 
d\xd^1\wedge d\xd^2 \wedge d\xd^3, 
\end{align}
and the ion number $4$-current $N_{\text{ion}}$ is prescribed as data. 

The measure on $\ca{E}_p$ in (\ref{electron_current}) is induced from
the $3$-form $\iota_X \# 1$,
\begin{align}
\notag
\iota_X \# 1 &= \sqrt{|\text{det}\mathfrak{g}|}^{\bm{V}} \frac{1}{3!}
\xd^a \epsilon_{abcd} dx^{b \bm{H}} \wedge dx^{c \bm{H}} \wedge dx^{d
  \bm{H}}\\
\label{i_X_hash_1_on_TM}
&\simeq
-\frac{\sqrt{|\text{det}{\mathfrak{g}}|}^{\bm{V}}}{g_{0c}^{\bm{V}}\xd^c}
dx^{1\bm{H}}\wedge dx^{2\bm{H}} \wedge dx^{3\bm{H}} 
\end{align}
where $\epsilon_{abcd}$ is the alternating symbol with
$\epsilon_{0123} = 1$.

The Vlasov-Maxwell equations constitute a non-linear
integro-differential system. Direct calculation of its solutions for
general plasma configurations is difficult and, to proceed
analytically, it is common to approximate the above as a finite number
of moments of $f$ in $\xd^a$ satisfying a non-linear field
system on $\ca{M}$ (a so-called
`fluid' model). However, there are difficult issues associated with
closing the resulting field system (see, for
example,~\cite{amendt:1986}) so we opt for a different approach. Our
strategy is to reduce the system by employing a discontinuous $f$, and
to proceed we need to cast (\ref{LVlas}) as an integral.    

One may rewrite (\ref{LVlas}) as
\begin{align}
 \label{dVlas}
& d(f\omega) \simeq 0,\\
\label{omega}
&\omega = \iota_L(\star 1^{\bm{V}} \wedge \iota_X\# 1) \in \Lambda_6
  T\ca{M}.
\end{align}
Integrating (\ref{dVlas}) over a 7-chain $\ca{A} \subset \ca{E}$ and
applying Stokes's theorem yields
\begin{equation}
 \int_{\partial \ca{A}} f\omega=0,  \label{intVlas}
\end{equation}
with $\partial \ca{A}$ the boundary of $\ca{A}$.  For differentiable
distributions, this equation is equivalent to (\ref{dVlas}); however,
since it makes no reference to the differentiability of $f$, it may be
regarded as a generalisation of (\ref{dVlas}) applicable to
discontinuous $f$. 

\section{Evolution of discontinuities}
Equation (\ref{intVlas}) may be used to develop an equation of motion
for a discontinuity, which we choose as a local hypersurface $\ca{H}$.
Suppose that $\ca{A}$ in (\ref{intVlas}) is a $7$-dimensional
 `pill-box' straddling $\ca{H}$. We may write
$\partial\ca{A} =
 \sigma_+ + \sigma_- + \sigma_0$ where $\sigma_+$ and $\sigma_-$ are the
 `top' and `bottom' of the pill-box and $\sigma_0$ is the `sides'
 of the pill-box. Thus, in the limit as the volume of $\ca{A}$ tends
 to zero with $\sigma_+$ tending to $\sigma$ and $\sigma_-$ tending
 to $-\sigma$, we recover the condition 
\begin{equation}
[f] \sigma^\ast \omega = 0,
\end{equation}
where the image of $\sigma$ is in $\ca{H}$ and $[f] = \sigma^*_+f +
\sigma^*_-f$. Thus it follows that a finite discontinuity in $f$ can occur
only across the image of a chain $\Sigma$ satisfying
\begin{equation}
 \Sigma^\ast \omega=0.  \label{Sigma_discon}
\end{equation}

Suppose that $\Sigma$ may be written locally
\begin{align}
\nonumber
\Sigma : \,\,\ca{V} \times \ca{D} &\rightarrow \ca{E}\subset T\ca{M}\\
(x^a,\xi^1,\xi^2) &\mapsto (x^a,\xd^b=\dot{\Sigma}^b(x,\xi))
\end{align}  
for $\ca{V}\subset\ca{M}$, where $\dot{\Sigma}^b$ denotes the $\xd^b$ component of $\Sigma$, and
$(\xi^1,\xi^2) \in \ca{D}\subset\mathbb{R}^2$. It is then
possible to translate (\ref{Sigma_discon}) into a field equation for a
family of vector fields $V_\xi$ on $\ca{V}$ given as  
\begin{equation}
 V_\xi (p) = V^a_\xi (x(p)) \partial_a = \dot{\Sigma}^a(x(p),\xi) \partial_a
\end{equation}
where, since $g_{ab}(x(p))\xd^a\xd^b=-1$ at $p\in\ca{E}$, it follows
\begin{equation}
 \label{norm}
g(V_\xi, V_\xi)=-1.
\end{equation}

Using (\ref{omega}, \ref{Sigma_discon}) it follows
\begin{equation}
 \Sigma^\ast(\underbrace{\iota_L \star 1^{\bm{V}} \wedge \iota_X \#
   1}_{(a)} + \underbrace{\star 1^{\bm{V}} \wedge \iota_L \iota_X \#
   1}_{(b)})=0.  \label{split}
\end{equation}

Consider first the term $(a)$ in equation~(\ref{split}):
\begin{equation}
\label{term_a}
 \Sigma^\ast (\iota_L \star 1^{\bm{V}} \wedge \iota_X \# 1)= \Sigma^\ast
 (\xd^{a} \iota_{\partial^{\bm{H}}_a} \star 1^{\bm{V}} \wedge \iota_X\# 1)
\end{equation}
where (\ref{Liou}) and $\iota_{{\bf f}^{\bm{V}}_a} \star 1^{\bm{V}}=0$
have been used (see (\ref{V_on_V}) in Appendix
\ref{appendix:vertical_and_horizontal}). Thus
\begin{equation}
\Sigma^\ast (\xd^{a} \iota_{\partial^{\bm{H}}_a} \star
1^{\bm{V}} \wedge \iota_X\# 1)
= \star \widetilde{V}_\xi \wedge \Sigma^\ast \iota_X \# 1,
\end{equation}
since $\iota_{\partial^{\bm{H}}_a} \star 1^{\bm{V}} = (g_{ab} \star
dx^b)^{\bm{V}}$ (see (\ref{V_on_H_and_H_on_V}) in Appendix
\ref{appendix:vertical_and_horizontal}).
Furthermore, using (\ref{H_lift_of_dx}) in Appendix
\ref{appendix:vertical_and_horizontal}, it follows
\begin{equation}
\label{pullback_dx_H}
\Sigma^\ast (dx^{a \bm{H}}) =
DV^a_\xi + \underline{d} \dot{\Sigma}^a
\end{equation}
where $D$ is the exterior covariant derivative on $\ca{M}$
and $\underline{d}$ is the exterior derivative on $\ca{D}$. Using
(\ref{i_X_hash_1_on_TM}, \ref{pullback_dx_H}) it follows
\begin{equation}
\label{Sigma_pullback_i_X_hash 1}
 \Sigma^\ast \iota_X \# 1 = \sqrt{|\text{det}\mathfrak{g}|}
 \frac{1}{3!} \epsilon_{abcd} V^a_\xi (DV^b_\xi +\underline{d}
 \dot{\Sigma}^b) \wedge (DV^c_\xi +\underline{d} \dot{\Sigma}^c)
 \wedge (DV^d_\xi +\underline{d} \dot{\Sigma}^d)
\end{equation}
and (\ref{term_a}) is
\begin{equation}
\label{simplified_term_a}
 \Sigma^\ast (\iota_L \star 1^{\bm{V}} \wedge \iota_X \# 1)= \star
 \widetilde{V}_\xi \wedge \frac{1}{2!} \sqrt{|\text{det}\mathfrak{g}|}
 \epsilon_{abcd} V^a_\xi DV^b_\xi \wedge \underline{d} \dot{\Sigma}^c
 \wedge \underline{d} \dot{\Sigma}^d
\end{equation}
since $\star \widetilde{V}_\xi \wedge DV^a_\xi \wedge DV^b_\xi=0$
($\text{dim}(\ca{M})=4$) and $\underline{d} \dot{\Sigma}^a \wedge
\underline{d} \dot{\Sigma}^b \wedge \underline{d} \dot{\Sigma}^c=0$
(dim($\ca{D}$)=2).

Since $\star \widetilde{V}_\xi \wedge DV^b_\xi= -(\nabla_{V_\xi}
V_\xi)^b \star 1$ and $\sqrt{|\text{det}\mathfrak{g}|}\epsilon_{abcd}=
\iota_{\partial_d}\iota_{\partial_c}\iota_{\partial_b}\iota_{\partial_a}\star
1$, it follows (\ref{simplified_term_a}) may be written
\begin{equation}
\label{term_a_on_M}
 \Sigma^\ast (\iota_L \star 1^{\bm{V}} \wedge \iota_X \# 1)=
 \widetilde{V}_\xi \wedge \nabla_{V_\xi} \widetilde{V}_\xi \wedge
 \Omega_\xi \wedge d\xi^1 \wedge d\xi^2,
\end{equation}
where the family of $2$-forms
$\Omega_\xi$ on $\ca{V}$ is
\begin{equation}
\label{surface_form}
 \Omega_\xi = \frac{\partial
 \dot{\Sigma}^a}{\partial \xi^1} \frac{\partial
 \dot{\Sigma}^b}{\partial \xi^2} g_{ac}\,g_{bd}\, dx^c \wedge dx^d.  
\end{equation} 

The second term $(b)$ in~(\ref{split}) can be rewritten using a similar
procedure :
\begin{equation}
 \Sigma^\ast (\star 1^{\bm{V}} \wedge \iota_L\iota_X \# 1) = \star 1
 \wedge \Sigma^\ast (\iota_L \iota_X \# 1)
\end{equation}
and from (\ref{Liou}, \ref{force}, \ref{i_X_hash_1_on_TM}) it follows
\begin{equation}
 \iota_L \iota_X \# 1 = -\frac{1}{2!}\qm\sqrt{|\det\mathfrak{g}|}^{\bm{V}} F^b{}^{\bm{V}}_e \xd^a \xd^e \epsilon_{abcd} dx^{c \bm{H}} \wedge dx^{d \bm{H}}.
\end{equation}
where (\ref{H_on_H}) and (\ref{V_on_H_and_H_on_V}) have been used.
Then
\begin{equation}
 \Sigma^\ast (\iota_L \iota_X \# 1)=
 -\frac{1}{2!}\qm\sqrt{|\text{det}\mathfrak{g}|} F^b{}_e V^a_\xi
 V^e_\xi \epsilon_{abcd} (DV^c_\xi +\underline{d} \dot{\Sigma}^c)
 \wedge (DV^d_\xi +\underline{d} \dot{\Sigma}^d)
\end{equation}
and
\begin{equation}
\label{term_b_on_M}
\Sigma^\ast ( \star 1^{\bm{V}} \wedge \iota_L \iota_X \# 1)= -\qm
\widetilde{V}_\xi \wedge \iota_{V_\xi} F \wedge \Omega_\xi \wedge d\xi^1
\wedge d\xi^2.
\end{equation}

Combining (\ref{term_a_on_M}, \ref{term_b_on_M}) and (\ref{split}) yields 
\begin{equation}
\label{discont_on_M}
 \Sigma^\ast \omega = \widetilde{V}_\xi \wedge ( \nabla_{V_\xi} \widetilde{V}_\xi - \qm \iota_{V_\xi} F) \wedge \Omega_\xi \wedge d\xi^1 \wedge d\xi^2 =0.
\end{equation}
Acting on (\ref{discont_on_M}) successively with $\iota_{V_\xi}$,
$\iota_{\partial/ \partial \xi^1}$ and $\iota_{\partial/ \partial
  \xi^2}$, and noting that $\iota_{V_\xi} \Omega_\xi=0$ and $g(V_\xi,V_\xi)=-1$, yields
\begin{equation}
 (\nabla_{V_\xi} \widetilde{V}_\xi - \qm \iota_{V_\xi} F) \wedge \Omega_\xi=0.  \label{discon}
\end{equation}
Thus, solutions to (\ref{discon}) may be obtained by demanding that
$V_\xi$ is driven by the Lorentz force 
\begin{equation}
\label{lorentz_force_law}
\nabla_{V_\xi} \widetilde{V}_\xi = \qm \iota_{V_\xi} F.
\end{equation}
However, although (\ref{lorentz_force_law}) is simpler than
(\ref{discon}), there are simple solutions to (\ref{discon}) that
do not satisfy (\ref{lorentz_force_law}); we will return to this point shortly.
\section{Non-linear electrostatic oscillations}
A laser pulse travelling through a plasma can excite plasma
oscillations, which induce very high longitudinal electric fields.
Due to nonlinear effects, there is a maximum amplitude of electric
field (the `wave-breaking limit') that can be sustained in the
plasma and, as mentioned in the introduction, our aim is to investigate the
relationship between the shape of the distribution (i.e. $\Sigma$) and the wave-breaking limit.

For simplicity, we choose to describe
the plasma using a distribution $f$ where $f=\alpha$ is a positive constant inside a
$7$-dimensional region $\mathcal{U}\subset\mathcal{E}$ and $f=0$ outside. In
particular, we consider $\mathcal{U}$ to be the union over each point
$p\in\ca{M}$ of a domain $\mathcal{W}_p$ whose boundary
$\partial\ca{W}_p$ in $\mathcal{E}_p$ is topologically equivalent to the
$2$-sphere. Such distributions are sometimes called `waterbags' in the
literature and can be completely characterized by $V_\xi$ and the
constant $\alpha$. Our approach may be considered
as a multi-dimensional generalization of the purely
1-dimensional relativistic waterbag model used
in~\cite{katsouleas:1988} to examine wave-breaking.

We work in Minkowski spacetime $(\ca{M},g)$ and assume that the ions
constitute a homogeneous immobile background. We employ an
inertial coordinate system $(x^a)$ adapted to the ions:
\begin{equation}
N_\text{ion}= n_\text{ion} \partial_0,
\end{equation}
where  
\begin{equation}
 g=-dx^0 \otimes dx^0 + dx^1 \otimes dx^1 + dx^2 \otimes dx^2 + dx^3 \otimes dx^3
\end{equation}
and the ion proper number density $n_\text{ion}$ is constant.

To proceed further we seek a form for $\Sigma$ axisymmetric
about $\dot{x}^3$ whose pointwise dependence in 
$\ca{M}$ is on the wave's phase $\zeta = x^3 - v x^0$ only, where $v$
is constant and $0<v<1$. We suppose that all electrons described by $f$ are travelling
slower than the wave, and the wave `breaks' if the
longitudinal velocity of any plasma electron equals $v$ (i.e. an
electron `catches up' with the wave). 

Introduce
\begin{equation}
 \label{coframe}
\bm{e}^1 = v dx^3 - dx^0,\qquad \bm{e}^2 = dx^3 - v dx^0,
\end{equation}
and decompose $\widetilde{V}_\xi$ as
\begin{equation}
\widetilde{V}_\xi = [\mu(\zeta) + A(\xi^1)]\, \bm{e}^1 + \psi(\xi^1,\zeta)\, \bm{e}^2
\,\,+ R\sin(\xi^1)\cos(\xi^2)dx^1 + R\sin(\xi^1)\sin(\xi^2)dx^2  \label{V_ansatz}
\end{equation}
for $0 < \xi^1 < \pi$, $0 \le \xi^2 < 2\pi$ 
where $R>0$ is constant.

Here, $(\gamma \bm{e}^1, \gamma \bm{e}^2, dx^1, dx^2)$, with $\gamma = 1/\sqrt{1-v^2}$, is an
orthonormal coframe on $\ca{M}$ adapted to $\zeta$.
Since $V_\xi$ is future-directed and timelike, and $\bm{e}^1$ is
timelike, it follows $\bm{e}^1(V_\xi) < 0$ and $\mu+A(\xi^1) > 0$.

The component $\psi$ is determined using (\ref{norm}),
\begin{equation}
\label{psi}
\psi = -\sqrt{[\mu + A(\xi^1)]^2 - \gamma^2[1 + R^2 \sin^2(\xi^1)]},
\end{equation}
where the negative square root is chosen because no electron is moving
faster along $x^3$ than the wave.

Substituting the {\em ansatz} (\ref{V_ansatz}) together with a purely
longitudinal electric field depending only on $\zeta$, 
\begin{equation}
 F=E(\zeta)\, dx^0 \wedge dx^3,
\end{equation}
into (\ref{lorentz_force_law}) yields
\begin{equation}
\label{F_dmu}
E= \frac{1}{\gamma^2} \frac{m}{q} \frac{d \mu}{d\zeta}.
\end{equation}
Equation (\ref{F_dmu}) is used to eliminate $E$ from Maxwell
equations~(\ref{Max}) and obtain a differential equation for $\mu$.

The electron number current is calculated using
(\ref{electron_current}): 
\begin{equation}
 N(p) = \alpha \bigg(\int_{\ca{W}_p} \xd^a \iota_X\# 1 \bigg) \partial_a
\end{equation}
and (\ref{Max},
\ref{V_ansatz}, \ref{psi}) yield
\begin{align}
\notag
\frac{1}{\gamma^2}\frac{d^2\mu}{d\zeta^2} = &-
  \frac{q^2}{m}n_{\text{ion}}\gamma^2\\
 \label{ODE_mu}
&- \frac{q^2}{m}2\pi R^2 \alpha \int\limits^\pi_0 \bigg([\mu +
    A(\xi^1)]^2
\,\,- \gamma^2[1 + R^2
    \sin^2(\xi^1)]\bigg)^{1/2}\sin(\xi^1)\,\cos(\xi^1)\, d\xi^1
\end{align}
and
\begin{equation}
\label{norm_A}
2\pi R^2 \int\limits^\pi_0
A(\xi^1)\,\sin(\xi^1)\,\cos(\xi^1)\,d\xi^1 = - \frac{n_\text{ion}\gamma^2\,v}{\alpha}
\end{equation}
where $\alpha$ is the value of $f$ inside $\mathcal{W}_p$.

The form of the 2nd order autonomous
non-linear differential equation (\ref{ODE_mu}) for $\mu$ is fixed by
specifying the generator $A(\xi^1)$ of $\partial\mathcal{W}_p$ subject to the
normalization condition (\ref{norm_A}).

\subsection{Electrostatic wave-breaking}
The form of the integrand in (\ref{ODE_mu}) ensures that the magnitude
of oscillatory solutions to (\ref{ODE_mu}) cannot be arbitrarily
large. For our model, the wave-breaking value $\mu_{\text{wb}}$
is the largest $\mu$ for which the argument of the square root in
(\ref{ODE_mu}) vanishes,
\begin{equation}
\mu_{\text{wb}} =
\text{max}\bigg\{-A(\xi^1) + \gamma\sqrt{1+R^2\sin^2(\xi^1)}
\,\bigg|\,0\le\xi^1\le\pi\bigg\},  \label{mu_wave-breaking}
\end{equation}
because $\mu<\mu_{\text{wb}}$ yields an imaginary integrand in
(\ref{ODE_mu}) for some
$\xi^1$. The positive square root in (\ref{mu_wave-breaking}) is
chosen because, as discussed above, $\mu + A(\xi^1) > 0$ and in
particular $\mu_\text{wb} + A(\xi^1)>0$.

The wave-breaking limit $E_{\text{max}}$ is obtained by evaluating
the first integral of (\ref{ODE_mu}) between $\mu_{\text{wb}}$ where
$E$ vanishes and the 
equilibrium\footnote{\label{footnote1}Note that the equilibrium of
  $\mu$ need not coincide with the plasma's thermodynamic equilibrium.} value
$\mu_{\text{eq}}$ of $\mu$ where $E$ is at a maximum. Using
(\ref{norm_A}) to eliminate $\alpha$ it follows that $\mu_{\text{eq}}$ satisfies
\begin{align}
\notag
\frac{1}{v}\int\limits^\pi_0 A(\xi^1)&\sin(\xi^1)\cos(\xi^1)\,d\xi^1\\
\label{mu_equilibrium}
&= \int\limits^\pi_0 \bigg([\mu_{\text{eq}} + A(\xi^1)]^2
- \gamma^2[1 + R^2
    \sin^2(\xi^1)]\bigg)^{1/2}
\sin(\xi^1)\cos(\xi^1)
    d\xi^1
\end{align}
with 
\begin{equation}
\label{A_negativity}
\int\limits^\pi_0 A(\xi^1)\sin(\xi^1)\cos(\xi^1)\,d\xi^1\, <\, 0
\end{equation}
since $\alpha, v >0$. Equation (\ref{ODE_mu}) yields the
maximum value $E_\text{max}$ of $E$,
\begin{align}
\notag E_{\text{max}}^2 = \,\,&2 m n_\text{ion}\Bigg[
-\mu_{\text{eq}} + \mu_{\text{wb}}
+ \, \frac{v}{\int\limits^\pi_0
    A(\xi^{1\prime})\sin(\xi^{1\prime})\cos(\xi^{1\prime})d\xi^{1\prime}} \times \\
& \int\limits^{\mu_{\text{eq}}}_{\mu_{\text{wb}}}\int\limits^\pi_0
\bigg([\mu + A(\xi^1)]^2
- \gamma^2 [1 + R^2
    \sin^2(\xi^1)]\bigg)^{1/2}
\sin(\xi^1)\cos(\xi^1)
    d\xi^1\,d\mu\Bigg].   \label{E_max}
\end{align}

The above is a general expression for $E_\text{max}$ given $A(\xi^1)$
as data. In the following, we determine a simple expression for an upper bound
on $E_\text{max}$ when $A(\xi^1) = -a\cos(\xi^1)$ where $a$ is a
positive constant ($a>0$ ensures (\ref{A_negativity}) is satified). Using (\ref{E_max}) it follows
\begin{align}
\label{E_max_simple}
E_{\text{max}}^2 = 2 m n_\text{ion}\Bigg\{
-\mu_{\text{eq}} + \mu_{\text{wb}}
+ \frac{3}{2}\frac{v}{a}
\int\limits^{\mu_{\text{eq}}}_{\mu_{\text{wb}}}
[\ca{I}_+(\mu) + \ca{I}_-(\mu)]\,d\mu
\Bigg\},
\end{align}
where
\begin{equation}
\label{I_plus_minus}
\ca{I}_\pm(\mu) = \pm\int\limits^1_0 \bigg([\mu \pm a\chi]^2
- \gamma^2[1 + R^2 (1-\chi^2)]\bigg)^{1/2}\chi\,d\chi
\end{equation}
Furthermore,
(\ref{mu_equilibrium}) may be written
\begin{equation}
\label{mu_equilibrium_simple}
\frac{3}{2}\frac{v}{a}[\ca{I}_+(\mu_{\text{eq}}) + \ca{I}_-(\mu_{\text{eq}})] = 1
\end{equation}
and since $\ca{I}_+(\mu_{\text{eq}})\ge \ca{I}_+(\mu)$ and 
$\ca{I}_-(\mu_\text{wb})\ge \ca{I}_-(\mu)$ for
$\mu_{\text{wb}}\le\mu\le\mu_{\text{eq}}$, using (\ref{E_max_simple},
\ref{mu_equilibrium_simple}) 
\begin{equation}
E^2_\text{max} \le \frac{3v}{a} m n_{\text{ion}}
(\mu_\text{eq}-\mu_{\text{wb}})[\ca{I}_-(\mu_{\text{wb}})-\ca{I}_-(\mu_{\text{eq}})]
\end{equation}
Furthermore $-\ca{I}_-(\mu_{\text{eq}}) \le
\frac{1}{2}\sqrt{\mu_\text{eq}^2 - \gamma^2}$ and $\ca{I}_-(\mu_{\text{wb}}) \le 0$
so
\begin{equation}
E^2_\text{max} \le \frac{3v}{2a} \frac{m^2\omega_p^2}{q^2} (\mu_\text{eq}- \mu_\text{wb})
\sqrt{\mu_\text{eq}^2 - \gamma^2}
\end{equation}
where $\omega_p=\sqrt{n_\text{ion}q^2/m}$ is the plasma
angular frequency (in units where $\varepsilon_0=1$ and $c=1$).
\subsection{Wave-breaking in an external magnetic field}
In tackling (\ref{discon}), one may opt to seek only those $V_\xi$
satisfying (\ref{lorentz_force_law}); this approach was followed
in the preceeding sections. Although, at first sight, this method appears to be a
simpler than attempting to solve (\ref{discon}), it is not always the
simplest option. There are potential advantages in considering
(\ref{discon}) in its generality, as we will now argue. 

The component of the magnetic field parallel to the velocity of a
point charge does not contribute to the Lorentz force on that point
charge. A similar observation may also be applied to certain $V_\xi$ in
(\ref{discon}) even though the $(\partial_1,\partial_2)$ components of $V_\xi$ are
non-zero. Furthermore, the results of the previous section are unaffected by a constant
magnetic field aligned along $x^3$.  

The axially symmetric $V_\xi$ introduced above is of the general form  
\begin{equation}
 V_\xi= (1+Y^2+ Z^2)^{1/2} \partial_0+ Y \cos(\xi^2) \partial_1+ Y
 \sin(\xi^2) \partial_2+ Z \partial_3,
\end{equation}
where $Y=\hat{Y}(x,\xi^1)$ and $Z=\hat{Z}(x,\xi^1)$. Suppose $F$ is of the form 
\begin{align}
&F = F_{\bm{I}} + F_{\bm{II}},\\
&F_{\bm{I}} = E(\zeta) dx^0\wedge dx^3,\\
&F_{\bm{II}} = B dx^1\wedge dx^2
\end{align}

We have
\begin{equation}
 i_{V_\xi} F_{\bm{II}}= B\, Y \big(\cos (\xi^2) dx^2 - \sin (\xi^2) dx^1 \big)
\end{equation}
and furthermore
\begin{align}
&\frac{\partial \dot{\Sigma}^1}{\partial \xi^2} = -Y\sin(\xi^2),\\
&\frac{\partial \dot{\Sigma}^2}{\partial \xi^2} = Y\cos(\xi^2)
\end{align}
so
\begin{equation}
g_{ab} \frac{\partial\dot{\Sigma}^a}{\partial \xi^2} dx^b =
-Y\sin(\xi^2)dx^1 + Y\cos(\xi^2)dx^2.
\end{equation}
Hence, using (\ref{surface_form}) it follows
\begin{equation}
\qm \iota_{V_\xi} F_{\bm{II}} \wedge \Omega_\xi=0
\end{equation}
and from (\ref{discon})
\begin{equation}
(\nabla_{V_\xi}\widetilde{V}_\xi - \qm \iota_{V_\xi} F_{\bm{I}})\wedge
  \Omega_\xi = 0.
\end{equation}
Therefore, if $V_\xi$ satisfies (\ref{discon}) with $F=F_{\bm{I}}$,
the same velocity field also satisfies (\ref{discon}) with
$F=F_{\bm{I}} + F_{\bm{II}}$. It follows that a longitudinal
magnetic field does not influence an axially symmetric discontinuity
in the electron distribution and the results of the previous section
hold for non-zero constant $B$.
\section{Beyond the Lorentz force}
The previous discussion clearly shows that there is merit in considering
(\ref{discon}) in its generality. We argue that for future extension
of this work to fields with more complicated spacetime dependence, it is prudent to eschew
(\ref{lorentz_force_law}) in favour of (\ref{discon}). We now
illustrate this point further using a very simple example.

Let the chain $\Sigma$ be such that
\begin{equation}
\label{constant_energy_V}
 V_\xi= (1+R^2)^{1/2} \partial_0+ R \sin(\xi^1)
 \cos(\xi^2) \partial_1 + R \sin(\xi^1) \sin(\xi^2)
 \partial_2 + R\cos(\xi^1) \partial_3
\end{equation}
where $R$ is a function on $\ca{M}$ and, using (\ref{surface_form}),
\begin{equation}
 \Omega_\xi= R^2 \sin(\xi^1) \cos(\xi^1) dx^1 \wedge dx^2+ R^2
 \sin^2(\xi^1) dx^3 \wedge \Big( \sin(\xi^2) dx^1- \cos(\xi^2) dx^2 \Big). 
\end{equation}

The $4$-acceleration of $V_\xi$ is
\begin{align}
\notag
\nabla_{V_\xi} \widetilde{V}_\xi = -&\frac{R\,
  V_\xi R}{\sqrt{1+R^2}}\,dx^0\\
\label{acceleration}
&+ V_\xi R \Big( \sin (\xi^1) \cos (\xi^2) dx^1+ \sin (\xi^1) \sin
(\xi^2) dx^2+ \cos (\xi^1) dx^3 \Big)
\end{align}
and, for simplicity, we assume that (\ref{acceleration}) is in response to a longitudinal electric field,
\begin{equation}
 F= E dx^0 \wedge dx^3,
\end{equation}
which contributes to (\ref{discon}) as
\begin{equation}
 \iota_{V_\xi} F= E \bigg( (1+R^2)^{1/2}dx^3 - R\cos (\xi^1) dx^0 \bigg).
\end{equation}

Clearly the Lorentz force equation (\ref{lorentz_force_law}) cannot be satisfied for general
$R$, since $\nabla_{V_\xi} \widetilde{V}_\xi$ contains terms in
$dx^1$, $dx^2$ which cannot cancel against terms in $\iota_{V_\xi} F$.
However, the only nonzero contribution to the left-hand side of
(\ref{discon}) can be made to vanish by requiring 
\begin{equation}
\label{VR}
 V_\xi R = \qm E \sqrt{1+R^2} \cos (\xi^1).
\end{equation}
Inspection of the $\xi$ dependences in (\ref{constant_energy_V}) and
(\ref{VR}) reveals that $R$ can depend only on $x^3$ and
\begin{equation}
 \frac{d}{dx^3} \sqrt{1+R^2}= \qm E
\end{equation}
so $E$ also depends only on $x^3$.

To compare the above with solutions to (\ref{lorentz_force_law}), we
seek a reparameterisation of $\Sigma$ whose corresponding family of
$4$-velocities satisfies (\ref{lorentz_force_law}). In particular, we
consider a map $\rho$
\begin{eqnarray}
\nonumber \rho : \ca{V} \times \ca{D}^\prime &\rightarrow& \ca{V} \times \ca{D}  \\
(x, \xi^{\prime 1}, \xi^{\prime 2}) &\mapsto& (x,
\xi^1=\psi^1(x,\xi^\prime),\xi^1=\psi^2(x,\xi^\prime))
\label{diff}
\end{eqnarray}
where $\ca{V}\subset\ca{M}$ and $(\xi^{\prime 1},\xi^{\prime 2})\in\ca{D}^\prime\subset\mathbb{R}^2$.
Then given $\Sigma$ satisfying $\Sigma^\ast \omega=0$,
\begin{equation}
\label{reparam_discont}
 (\Sigma \circ \rho)^\ast \omega= \rho^\ast(\Sigma^\ast \omega)=0.
\end{equation}
The chains $\Sigma$ and $(\Sigma \circ \rho)$ locally represent the same
discontinuity, and are physically equivalent. However, the families $V_\xi= V^a_\xi
\partial_a$ and $W_{\xi^\prime}=W^a_{\xi^\prime} \partial_a$ of vector
fields, where 
\begin{equation}
 V^a_\xi = \Sigma^\ast \xd^a, \qquad W^a_{\xi^\prime} = (\Sigma \circ \rho)^\ast \xd^a,
\end{equation}
are different. We demand
\begin{equation}
 \nabla_{W_\xi^\prime} \widetilde{W}_{\xi^\prime} = \qm \iota_{W_{\xi^\prime}} F
\end{equation}
and using (\ref{reparam_discont}) it follows
\begin{equation}
 W_{\xi^\prime}\psi^2 =0, \qquad W_{\xi^\prime}\psi^1 = -\frac{q E}{mR}
 \sqrt{1+R^2}\sin (\psi^1).  \label{thetaeq}
\end{equation}
One may solve (\ref{thetaeq}) to determine $(\Sigma\circ\rho)$, but is
clear that (for general $E$) solving for the discontinuity in terms of
$\Sigma$ is a simpler task. Furthermore, we expect this state of affairs to hold for
more complicated configurations. 
\section{Conclusion}
We have developed a covariant formalism for tackling discontinuities
in $1$-particle distributions. We have used it to develop 
wave-breaking limits for models of thermal plasmas whose distributions have effectively $1$-dimensional spacetime
dependence but are $3$-dimensional in velocity. 
\appendix 

\section{Vertical and horizontal lifts}
\label{appendix:vertical_and_horizontal}
Given tensors on a manifold $\ca{M}$, there are a number of ways to lift
them onto the tangent manifold $T\ca{M}$.  The simplest and best known of
these are the vertical lift and (given a connection on $\ca{M}$) the
horizontal lift.  The following is a summary of some important
properties of these lifts; for more details see, for example,~\cite{yano:1973}.

The vertical lift is a tensor homomorphism.  Acting on forms, it is equivalent to the pull-back with respect to the canonical projection map $\pi : T\ca{M} \rightarrow \ca{M}$:
\begin{equation}
 \beta^{\bm{V}}= \pi^\ast \beta, \qquad \forall\,\, \beta \in \Lambda \ca{M}.
\end{equation}
Acting on a vector $Y$, the vertical lift is
\begin{equation}
 Y^{\bm{V}}= Y^a \frac{\partial}{\partial \xd^a} \qquad
 \forall\,\,Y=Y^a \frac{\partial}{\partial x^a} \in T\ca{M}.
\end{equation}

Note from these definitions, contraction of the vertical lift of a vector with the vertical lift of a form vanishes:
\begin{equation}
\label{V_on_V}
 \iota_{(Y^{\bm{V}})} \beta^{\bm{V}}=0.
\end{equation}

The horizontal lift makes use of a connection $\nabla$, with
coefficients $\Gamma^a{}_{bc}$ in a coordinate basis $\partial_a = \partial/\partial x^a$:
\begin{equation}
 \nabla_{\partial_c} \partial_b = \Gamma^a{}_{bc} \partial_a.
\end{equation}
The horizontal lift $dx^{a \bm{H}}$ of the basis form $dx^a$ is
\begin{equation}
\label{H_lift_of_dx}
 dx^{a \bm{H}}= d\xd^a + \xd^c \Gamma^a{}^{\bm{V}}_{bc} dx^b,
\end{equation}
while that of the basis vector $\partial/\partial x^a$ is
\begin{equation}
 \bigg( \frac{\partial}{\partial x^a}\bigg)^{\bm{H}}= \frac{\partial}{\partial x^a} - \xd^c \Gamma^b{}^{\bm{V}}_{ac} \frac{\partial}{\partial \xd^b}.
\end{equation}
The horizontal lifts of more general 1-forms and vectors may be determined from the relations
\begin{align}
 & (f\beta)^{\bm{H}}= f^{\bm{V}} \beta^{\bm{H}} \qquad \beta \in \Lambda^1 \ca{M},\\
& (fY)^{\bm{H}}= f^{\bm{V}} Y^{\bm{H}} \qquad \forall\,\, Y\in T\ca{M},
\end{align}
for any function $f$ on $\ca{M}$.

Similarly to the vertical lift, the contraction of the horizontal lift of a vector with the horizontal lift of a form vanishes:
\begin{equation}
\label{H_on_H}
 \iota_{(Y^{\bm{H}})} \beta^{\bm{H}}=0.
\end{equation}

Two other useful identities relate to the contractions of vertical and horizontal lifts of forms and vectors:
\begin{equation}
\label{V_on_H_and_H_on_V}
 \iota_{(Y^{\bm{V}})} \beta^{\bm{H}}=\iota_{(Y^{\bm{H}})}
 \beta^{\bm{V}}=(\iota_Y \beta)^{\bm{V}}.
\end{equation}
\acknowledgments
We thank RA Cairns, B Ersfeld, J Gratus, AJW Reitsma and RMGM Trines
for useful discussions. This work is supported by EPSRC grant
EP/E022995/1 and the Cockcroft Institute.  

\end{document}